\documentclass[10pt]{sig-alternate}
\usepackage{multirow}
\usepackage{array}
\usepackage{enumitem}
\usepackage{subcaption}
\usepackage{tikz}
\usepackage{pgfplots}
\usepackage[utf8]{inputenc}
\usepackage{graphicx}

\setlist[enumerate,1]{%
  label=\arabic*.,
}

\newlist{inlinelist}{enumerate*}{1}
\setlist*[inlinelist,1]{%
  label=(\roman*),
}

\begin{document}

\title{Rama: Controller Fault Tolerance in \\ Software-Defined Networking Made Practical}
\author{Andr{\'e} Mantas, Fernando M. V. Ramos\\
\affaddr{LASIGE, Faculdade de Ci\^{e}ncias, Universidade de Lisboa, Portugal}\\
\email{amantas@lasige.di.fc.ul.pt, fvramos@ciencias.ulisboa.pt}\\
}

\maketitle

\begin{abstract}
In Software-Defined Networking (SDN), network applications use the logically centralized network view provided by the controller to remotely orchestrate the network switches.
To avoid the controller being a single point of failure, traditional fault-tolerance techniques are employed to guarantee availability, a fundamental requirement in production environments.
Unfortunately, these techniques fall short of ensuring correct network behaviour under controller failures.
The problem of these techniques is that they deal with only part of the problem: guaranteeing that application and controller state remains consistent between replicas.
However, in an SDN the switches maintain hard state that must also be handled consistently.
Fault-tolerant SDN must therefore include switch state into the problem.

A recently proposed fault-tolerant controller platform, Ravana, solves this problem by extending fault-tolerant SDN control with mechanisms that guarantee control messages to be processed transactionally and exactly once, at both the controllers and the switches.
These guarantees are given even in the face of controller and switch crashes.
The elegance of this solution comes at a cost.
Ravana requires switches to be modified and OpenFlow to be extended with hitherto unforeseen additions to the protocol.
In face of this challenge we propose Rama, a fault-tolerant SDN controller platform that offers the same strong guarantees as Ravana \emph{without} requiring modifications to switches or to the OpenFlow protocol.
Experiments with our prototype implementation show the additional overhead to be modest, making Rama the first fault-tolerant SDN solution that can be immediately deployable.
\end{abstract}

%--------------------------------------------------------------------------------------------
\section{Introduction}
\label{intro}
%--------------------------------------------------------------------------------------------
Software-Defined Networking (SDN) decouples the network control plane from the data plane via a well-defined programming interface (such as OpenFlow~\cite{mckeown2008}).
This decoupling allows the control logic to be logically centralized, easing the implementation of network policies, enabling advanced forms of traffic engineering (e.g., \\Google's B4~\cite{jain2013}), and facilitating innovation (network virtualization~\cite{koponen2014} being a prominent example).

\begin{figure}[h!]
    \centering
	\includegraphics[width=0.45\textwidth]{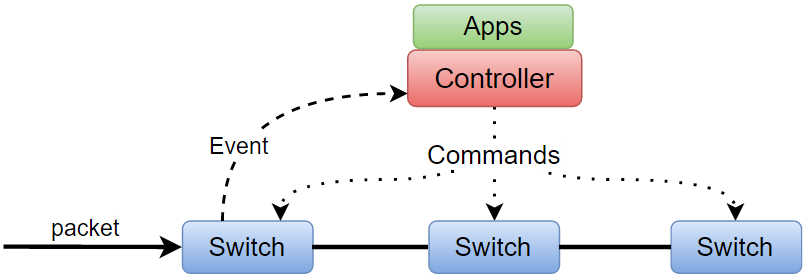}
    \caption[SDN flow execution]{SDN flow execution. Switches send events to the controller as needed and the controller replies with one or more commands that modify the switch's tables.}
    \label{fig:intro_sdn}
\end{figure}

The controllers are the crucial enabler of the SDN paradigm: they maintain the logically centralized network state to be used by applications and act as a common intermediary with the data plane.
Figure \ref{fig:intro_sdn} shows the normal execution in an SDN environment.
Upon receiving a packet it does not know how to handle, the switch sends an \emph{event} to the controller.
The controller delivers the event to the applications, which afterwards apply their logic based on this event, and eventually instruct the controller to send \emph{commands} to the switches (e.g., to install flow rules).

A trivial implementation of SDN using a centralized controller would lead to an undesirable outcome: a single point of failure.
To guarantee the required availability of network control, it is necessary the controller platform to be made fault-tolerant.
Fault tolerance demands transparency: for a controller that claims having such ability -- in other words, for it to be logically centralised -- applications that run on it should operate correctly in the presence of faults.
This is a fundamental requirement, as in case of controller failures the network view needs to be maintained consistent, otherwise applications will operate in a stale network view, leading to network anomalies that can have undesirable consequences (e.g., security breaches)~\cite{levin2012, katta2015}.

To address this problem, traditional replication techniques are usually employed~\cite{oki1988, lamport1998, ongaro2014}.
However, building a consistent network view in the controllers is not enough to offer consistent logically centralized control that prevents the above-mentioned anomalies.
In SDN, it is necessary to include switch state into the system model to achieve this goal~\cite{katta2015}.
Since switches are programmed by controllers (and controllers can fail), there must be mechanisms to ensure that the entire event-processing cycle of SDN is handled consistently.

A correct, fault-tolerant SDN environment needs to ensure \emph{observational indistinguishability}~\cite{katta2015} between an ideal central controller and a replicated controller platform.
Informally, to ensure observational indistinguishability the fault-tolerant system should behave the same way as a fault-free SDN for its users (end-hosts and network applications).
Fot this purpose, it is necessary the following properties to be met:

\begin{itemize}
\item \textbf{Total Event Ordering:} Controller replicas should process events in the same order and subsequently all controller application instances should reach the same internal state
\item \textbf{Exactly-Once Event Processing:} All the events are processed, and neither are lost nor processed repeatedly.
\item \textbf{Exactly-Once Execution of Commands:} Any given series of commands are executed once, and only once on the switches.
\end{itemize}

To the best of our knowledge, the problem of correct, fault-tolerant SDN control has only been addressed in the work by Katta~\emph{et al.}~\cite{katta2015}.
Instead of just keeping the controller state consistent, the authors propose Ravana, a fault-tolerant SDN controller platform that handles the entire event-processing cycle as a transaction -- either all or none of the components of this transaction are executed.
This enables Ravana to correctly handle switch state and thus guarantee SDN correctness even under fault.

To achieve these properties, however, Ravana requires modifications to the OpenFlow protocol and to existing switches.
Specifically, switches need to maintain two buffers, one for events and one for commands, and four new messages need to be added to the protocol.
These modifications preclude the adoption of Ravana on existing systems and hinder the possibility of it being used in the near future (as there are no guarantees these messages be added to OpenFlow anytime soon, for instance).

Faced with this challenge, we propose Rama, a fault-tolerant SDN controller platform that, similar to Ravana, offers a transparent control plane that allows unmodified network applications to run in a consistent and fault-tolerant environment.
The novelty of the solution lies in Rama not requiring changes to OpenFlow nor to the underlying hardware, allowing immediate deployment.
For this purpose, Rama exploits existing mechanisms in OpenFlow and orchestrates them to achieve its goals.

The main contributions of this work can be summarized as follows:

\begin{itemize}
\item A protocol for fault-tolerant SDN that provides the correctness  guarantees of a logically centralised controller \emph{without} requiring changes to OpenFlow or modifications to switches.
\item The implementation and evaluation of a prototype controller -- Rama -- that demonstrates the overhead of the solution to be modest.
\end{itemize}

%--------------------------------------------------------------------------------------------
\section{Fault tolerance in SDN}
\label{motivation}
%--------------------------------------------------------------------------------------------

Katta~\emph{et al.} have experimentally shown~\cite{katta2015} that traditional techniques for replicating controllers do not ensure correct network behaviour in case of failures.
The reason is that these techniques address only part of the problem: maintaining consistent state in controller replicas.
By not considering switch state (and the interaction controller-switches) inconsistencies may arise, resulting in potentially severe network anomalies.
In this section we present a summary of the problems of using techniques that do not incorporate switches in the system model, which lead to the design requirements of a \emph{correct} fault-tolerant SDN solution. 
We also present Ravana~\cite{katta2015}, the first fault-tolerant controller that achieves the required correctness guarantees for SDN.

\subsection{Inconsistent event ordering}

Since OpenFlow 1.3, switches can maintain TCP connections with multiple controllers.
In a fault-tolerant configuration switches can be set to send all their events to all known controller replicas.
As replicas process events as they are received, each one may end up building a different internal state.
Although TCP guarantees the order of events delivered by each switch, there are no ordering guarantees between events sent to controllers by the different switches, leading to the problem.

Consider a simple scenario with two controller replicas (c1 and c2) and two switches (s1 and s2) that send all events to both controllers.
Switch s1 sends two events -- e1 and e2, in this order -- and switch s2 sends two other events --  e3 and e4, in this order.
One possible outcome where both controllers receive events in a different order while respecting the TCP FIFO property is c1 receiving events in the order e1, e3, e2, e4 and c2 receiving in the order e3, e4, e1, e2.
Unfortunately, an inconsistent ordering of events can lead to incorrect packet-processing decisions~\cite{katta2015}.
As a result of this consistency problem we derive the first design goal for a fault-tolerant and correct SDN controller:

%%%
\textbf{Total event ordering:} controllers replicas should process the same (total) order of events and subsequently all controller application instances should reach the same internal state.
%%%%

\subsection{Unreliable event delivery}

In order to achieve a total ordering of events between controller replicas two approaches can be used:

\begin{enumerate}
\item The master (primary) replica can store controller state (including state from network applications) in an external consistent data-store (as in Onix \cite{koponen2010}, ONOS \cite{berde2014}, and SMaRtLight \cite{botelho2014});
\item The controller state can be kept consistent using replicated state machine protocols.
\end{enumerate}

Although both approaches ensure a consistent ordering of events between controller replicas, they are not fault-tolerant in the standard case where only the master controller receives all events.

If we consider –- for the first approach -- that the master replica can fail between receiving an event and finishing persisting the controller state in the external data-store (which happens after processing the event through controller applications), that event will be lost and the new master (i.e., one of the other controller replicas) will never receive it.
The same can happen in the second approach: the master replica can fail right after receiving the event and before replicating it in the shared log (which in this case happens before processing the event through the controller applications).
In these cases, since only the crashed master received the event, the other controller replicas will not have an updated view of the network.
Again, this may cause severe network problems~\cite{katta2015}. 
Similar problems can occur in case of repetition of events.
These problems lead to the second design goal:

%%%
\textbf{Exactly-once event processing:} All the events sent by switches are processed, and are neither lost nor processed repeatedly.
%%%

\subsection{Repetition of commands}

In either traditional state machine replication or consistent storage approaches, if the master controller fails while sending a series of commands, the new elected master may send repeated commands.
This may happen when the old master fails before informing the slave replica of its progress.
Since some commands are not idempotent~\cite{katta2015}, its duplication can lead to undesirable network behaviour. 
This problem leads to the third and final design goal:
%%%

\textbf{Exactly-once command execution:} any series of commands are executed only once on the switches.

\subsection{Ravana}

Ravana~\cite{katta2015} is the first controller to provide correct fault-tolerant SDN control.
To achieve this, Ravana processes control messages transactionally and exactly once (at both the controllers and the switches) using a replicated state machine approach, but without involving the switches in an expensive consensus protocol.

\begin{figure}[h]
  \centering
  \includegraphics[width=0.45\textwidth]{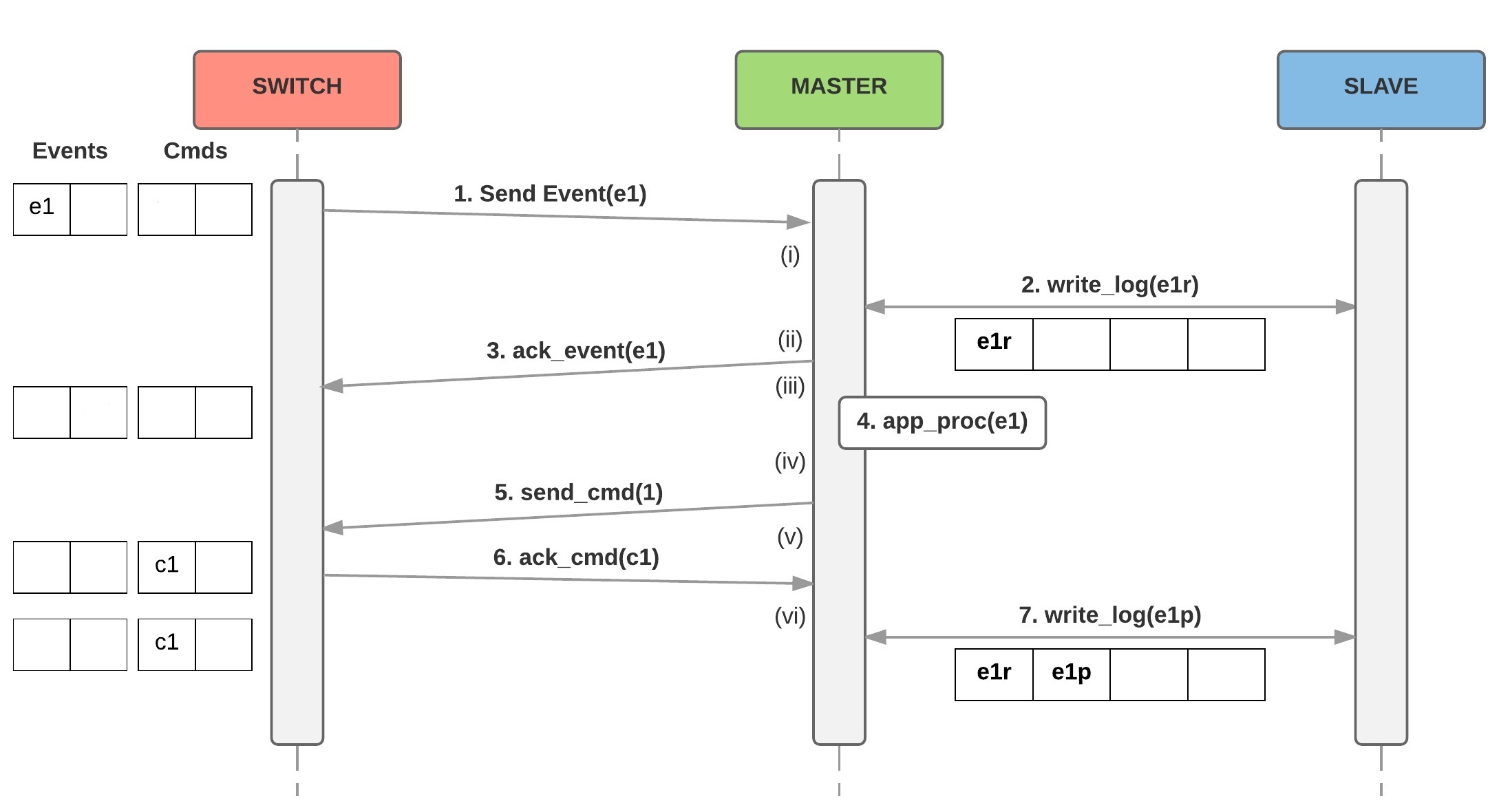}
  \caption[Ravana Protocol]{Ravana protocol. In Ravana switches maintain two buffers (displayed on the left) to re-transmit events and filter repeated commands in case of master failure. New acknowledge messages (\texttt{ack\_event} and \texttt{ack\_cmd}) are exchanged between the switch and the master to guarantee the consistency requirements.}
  \label{fig:ravana-protocol}
\end{figure}

The protocol used by Ravana is show in Figure \ref{fig:ravana-protocol}.
Switches buffer events (as they may need to be retransmitted) and send them to the master controller that will replicate them in a shared log with the slaves.
The controller will then reply back to the switch acknowledging the reception of the events.
Then, events are delivered to applications that may after processing require one or more commands to be sent to switches.
Switches reply back to acknowledge the reception of these commands and buffer them to filter possible duplicates.

While Ravana allows unmodified applications to run in a fault-tolerant environment, it requires modifications to the OpenFlow protocol and to switch hardware.
Namely, Ravana leverages on buffers implemented on switches to retransmit events and filter possible repeated commands received from the controllers.
Also, explicit acknowledgement messages must be added to the OpenFlow protocol so that the switch and the controller acknowledge received messages.
Unfortunately, these requirements preclude immediate adoption of Ravana.
For instance, it is not antecipated OpenFlow to be extended to include the required messages anytime soon.
These limitations are the main motivation for our proposal, which we present next.

%--------------------------------------------------------------------------------------------
\section{Rama design}
\label{design}
%--------------------------------------------------------------------------------------------

The goal of our work is to build a strongly consistent and fault-tolerant control plane for SDN to be used transparently by unmodified applications.
This section describes the architecture and protocol for such control plane, which is driven by the following four requirements. First, reliability: the system should maintain a correct and consistent state even in the presence of failures (in both the controllers and switches). Second, transparency: the consistency and fault-tolerance properties should be completely transparent to applications. Third, performance: the performance of the system should not degrade as the number of network elements (events and switches) grows. Fourth, immediate deployability: the solution should work with existing switches and not require new additions to the OpenFlow protocol.

\subsection{Architecture}

\begin{figure}[t]
    \includegraphics[width=0.2\textwidth]{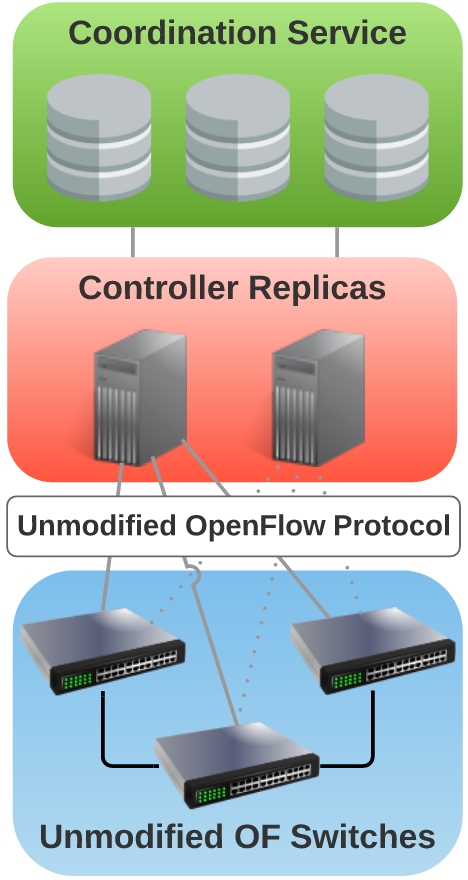}
    \centering
    \caption[High level architecture of the system]{High level architecture of the system.}
    \label{fig:arquitectura_alto_nivel}
\end{figure}

The high-level architecture of our system, Rama\footnote{In the Hindu epic Ramayana, Rama is the hero whose wife (Sita) is abducted by Ravana.}, is depicted in Figure \ref{fig:arquitectura_alto_nivel}.
Its main components are: (i) OpenFlow enabled switches (switches that are implemented according to the OpenFlow specification), (ii) controllers that manage the switches and (iii) a coordination service.
In our model, we consider only one network domain with one primary controller and one or more backup controllers, depending on the number of faults to tolerate.
Each switch connects to one primary controller and multiple (\textit{f} to be precise) backup controllers (to tolerate up to \emph{f} crash controller faults).
This primary/backup model is supported by OpenFlow in the form of master/slave and allows the system to tolerate controller faults.
When the master controller fails, the remaining controllers will elect a new leader to act as the new master for the switches managed by the crashed master.
This election is supported by the coordination service.

The coordination service offers strong consistency and abstracts controllers from complex primitives like fault detection and total order, making them simpler and more robust.
Note that the coordination system requires a number of replicas equal to \textit{2f+1}, with \textit{f} being the number of faults to tolerate.
The strong consistency model assures that updates to the coordination service made by the master will only return when they are persistently stored.
This means that slaves will always have the fresh modifications available as soon as the master receives confirmation of the update.
This results in a consistent network view among all controllers even if some fail.
The need for agreement between several replicas make the coordination service the system bottleneck~\cite{hunt2010}.
In addition to the controllers' state, the switches also maintain state that must be handled consistently in the presence of faults.
Fulfilling this request is the main goal of the protocol we present next.

\subsection{Rama protocol}
\label{sec:ft_protocol}

In an SDN setting, switches generate events (e.g., when they receive packets or when the status of a port changes) that are forwarded to controllers.
The controllers run multiple applications that process the received events and may send commands to one or more switches in reply to each event.
This cycle repeats itself in multiple switches across the network as needed.

\begin{figure}[h]
    \includegraphics[width=0.45\textwidth]{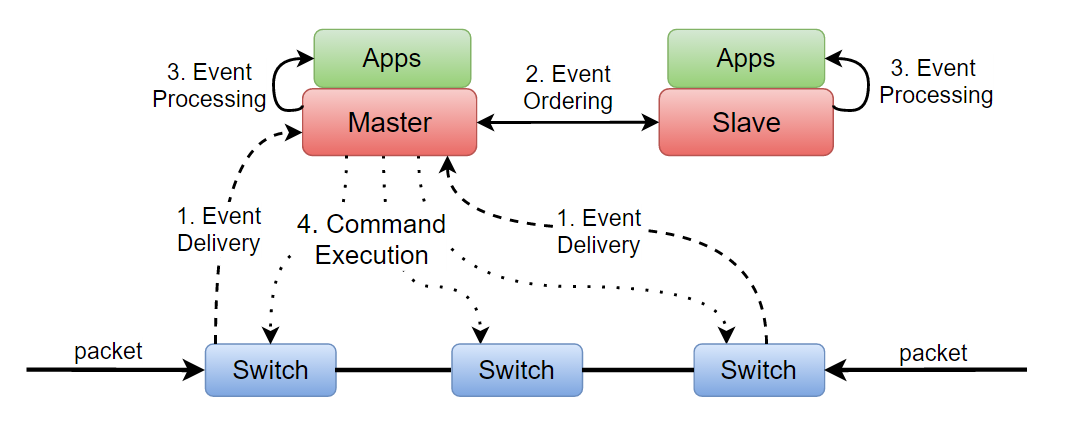}
    \centering
    \caption[Control loop]{Control loop of (1) event delivery, (2) event ordering, (3) event processing, and (4) command execution. Events are delivered to the master controller, which decides a total order on the received events. The events are processed by applications in the same order in all controllers. Applications issue commands to be executed in the switches.}
    \label{fig:control_loop}
\end{figure}

In order to maintain a correct system in the presence of faults, one must handle the state in the controllers and the state in the switches consistently.
To ensure this, the entire cycle presented in Figure \ref{fig:control_loop} is processed as a \emph{transaction}: either all or none of the components of this transaction are executed.
This means that
\begin{inlinelist}
\item the events are processed exactly once at the controllers,
\item all controllers process events in the same (total) order to reach the same state, and
\item the commands are processed exactly once in the switches.
\end{inlinelist}
Because the standard operation in OpenFlow switches is to simply process commands as they are received, the controllers must coordinate to guarantee the required exactly-once semantics.
Ravana~\cite{katta2015} does not need this coordination because the (modified) switches can simply buffer the commands received and discard repeated commands (i.e., those with the same identifier) sent by the new controller.

By default, in OpenFlow a master controller receives all asynchronous messages (e.g., \texttt{OFPT{\_}PACKET{\_}IN}), whe\-re\-as the slaves controllers only receive a subset (e.g., port modifications).
With this configuration only the master controller would receive the events generated by switches.
There are two options to solve this problem.
One is for slaves to change this behaviour by sending an \texttt{OFPT{\_}SET{\_}ASYNC} message to each switch that modifies the asynchronous configuration.
As a result, switches send all required events to the slaves.
Alternatively, all controllers can set their role to \texttt{EQUAL}.
The OpenFlow protocol specifies that switches should send all events to every controller with this role.
Then, controllers need to coordinate between themselves who the master is (i.e., the one that processes and sends commands).
We have opted for the second solution and use the coordination service for leader election amongst controllers.

\begin{figure}[ht!]
    \includegraphics[width=0.45\textwidth]{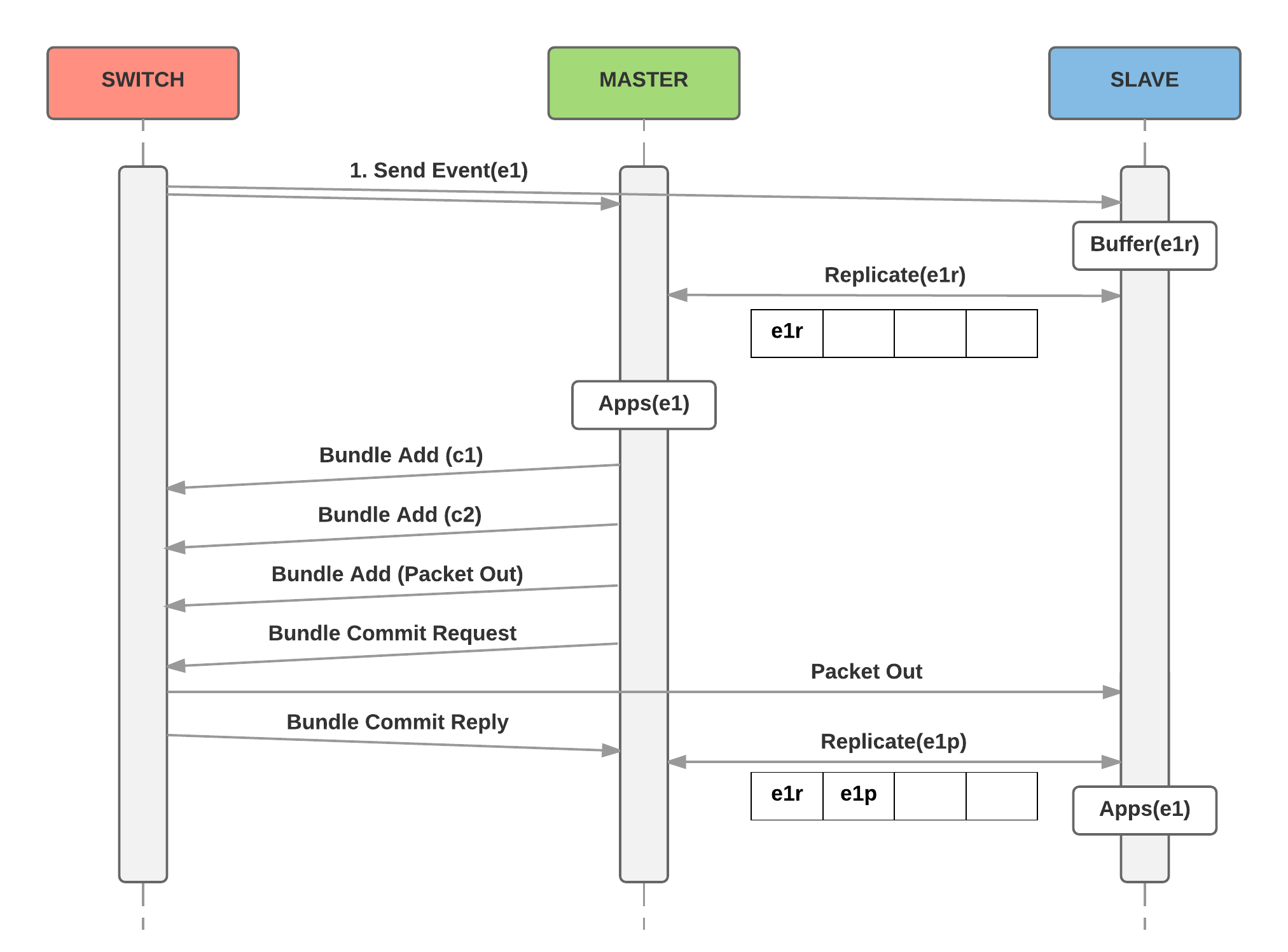}
    \centering
    \caption[Fault-free case of the protocol]{Fault-free case of the protocol. Switches send generated events to all controllers so that no event is lost. The master controller replicates the event in the shared log and then feeds its applications with the events in log order. Commands sent are buffered by the switch until the controller sends a Commit Request. The corresponding Commit Reply message is forwarded to all controllers.}
    \label{fig:rama-protocol}
\end{figure}

The fault-free execution of the protocol is represented in Figure \ref{fig:rama-protocol}.
In the figure we consider a switch to be connected with one master controller and a single slave controller.
The main idea is that switches must send messages to \textit{all controllers}, so that they can coordinate themselves even if some fail at any given point.
In Ravana, because switches simply buffer events (so that they can be retransmitted to a new master if needed), switches can send events only to the current master, instead of to every controller.

\begin{figure}[ht!]
    \includegraphics[width=0.45\textwidth]{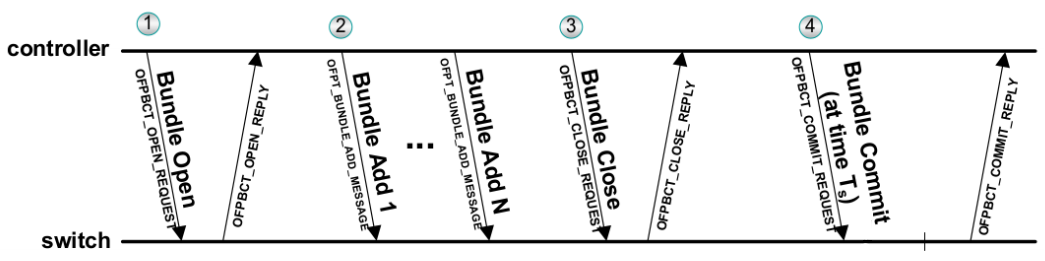}
    \centering
    \caption[OpenFlow Bundles]{OpenFlow Bundles.}
    \label{fig:openflow-bundles}
\end{figure}

The master controller then replicates the event in a shared log with the other controllers, imposing a total order on the events received (to simplify, the coordination service is omitted from the figure).
When the event is replicated to the shared log controllers, it is processed by the master controller applications, which will generate zero or more commands.
To guarantee exactly-once semantics, the commands are sent to the switches in bundles (a feature introduced in OpenFlow 1.4, see Figure \ref{fig:openflow-bundles}).
With this feature a controller can open a bundle, add multiple commands to it and then instruct the switch to commit all commands present in the bundle in an atomic and ordered fashion.

Rama uses bundles in the following way.
When an event is processed by all modules, the required commands are added by the master controller to a bundle.
The master then sends an \texttt{OFPBCT\_COMMIT\_REQUEST} message to each switch affected by the event.
The switch processes the request and tries to apply all the commands in the bundle in order.
Afterwards, it then sends a reply message indicating if the Commit Request was successful or not.
This message is used by Rama as an acknowledgement.

Again, we need to make sure that this reply message is sent to all controllers.
This is a challenge, because Bundle Replies are Controller-to-Switch messages and hence are only sent to the controller that made the request (using the same TCP connection).
To overcome this challenge we introduce a new mechanism in Rama.
The way we inform other controllers if the bundle was committed or not (so that they can decide later if they need to resend specific commands) is by including one \texttt{OFPT{\_}PACKET{\_}OUT} message in the end of the bundle with the action \texttt{output=controller}.
The outcome is that the switch will send the information included in the \texttt{OFPT{\_}PACKET{\_}OUT} message to all connected controllers in a \texttt{OFPT{\_}PACKET{\_}IN} message.
This message is set by the master controller to inform slave controllers about the events that were fully processed by the switch (in this bundle).
This prevents a new master from sending repeated commands, thus guaranteeing exactly-once semantics.
Ravana does not need to rely on bundles since switches buffer all received commands so that they can discard possible duplicates from a new master.

The master finishes the transaction by replicating an \texttt{event} \texttt{processed} \texttt{message} in the log, informing backup controllers that they can safely feed the corresponding event in the log to their applications.
This is done to simply bring the slaves to the same updated state as the master controller (the resulting commands sent by the applications are naturally discarded).

\subsubsection{Fault cases}
\label{subsec:fault-cases}

\begin{figure}[h]
    \includegraphics[width=0.45\textwidth]{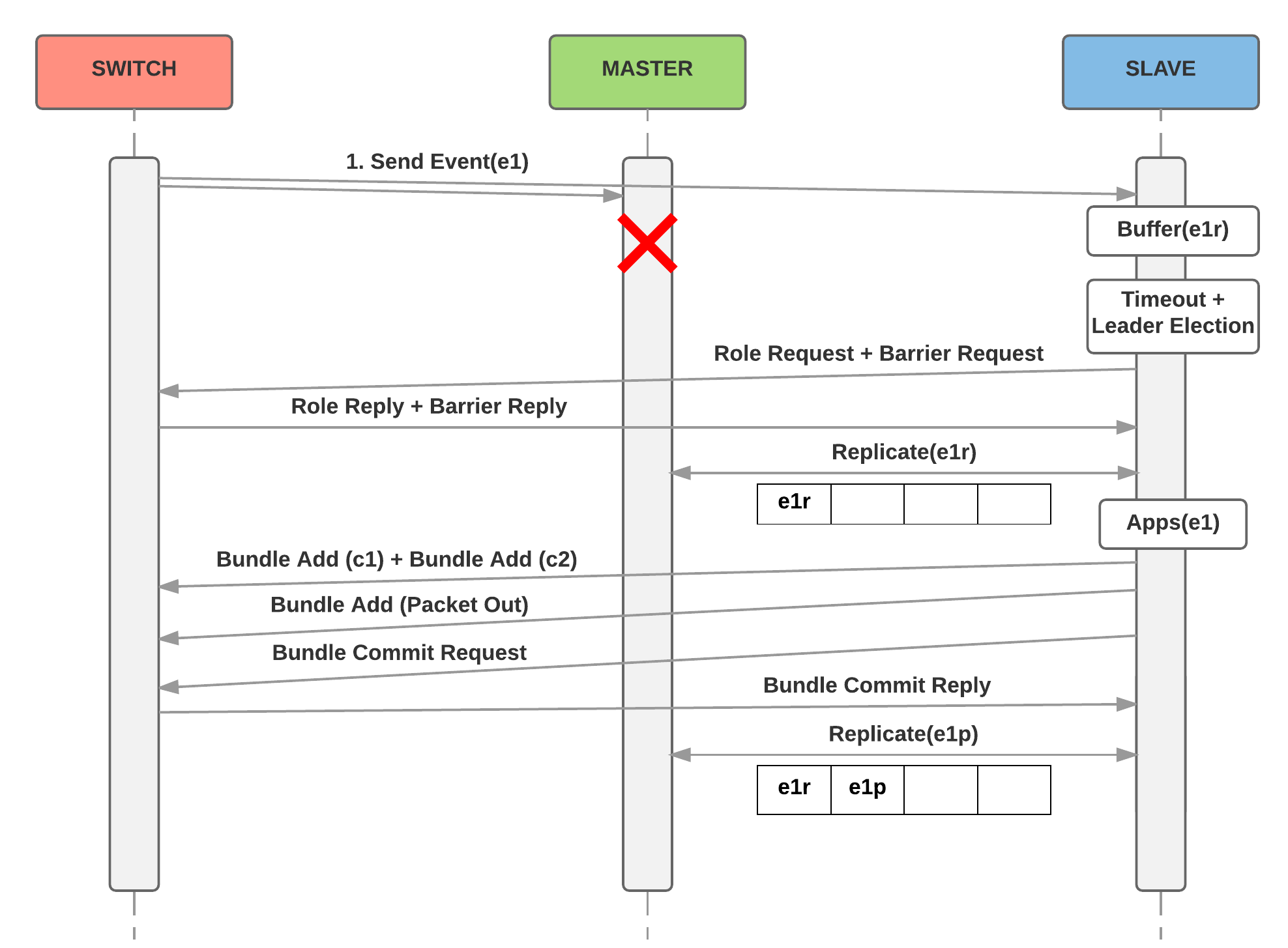}
    \centering
    \caption[Failure case 1 of the protocol]{Case of the protocol where the master fails before replicating the event received. Because the slaves buffer all events, the event is not lost and the new master can resume the execution of the failed controller.}
    \label{fig:rama-protocol-fault-case-1}
\end{figure}

\begin{figure}[ht]
    \includegraphics[width=0.45\textwidth]{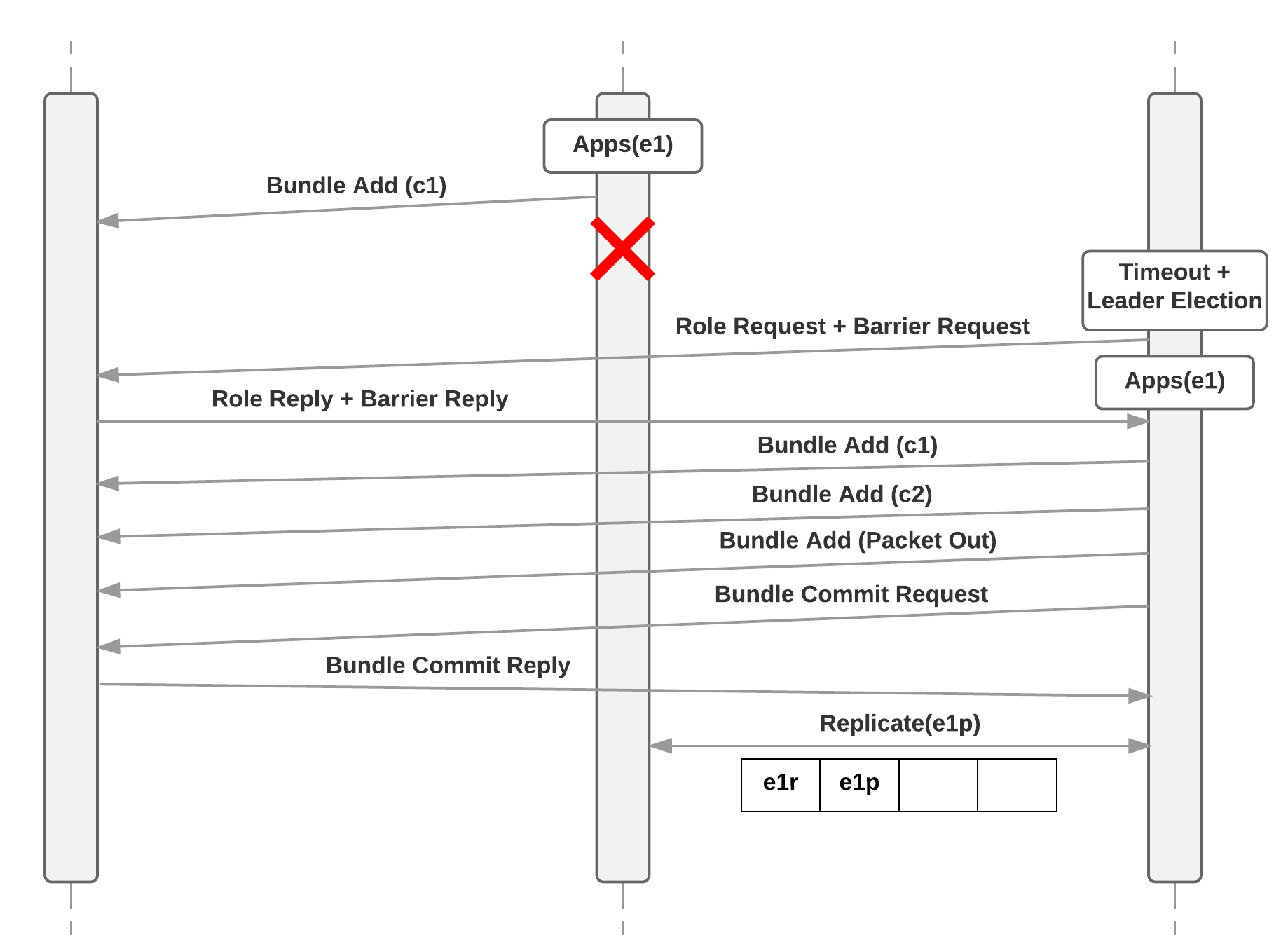}
    \centering
    \caption[Failure case 2 of the protocol]{Case of the protocol where the master fails after replicating the event. The first part of the protocol is identical to the fault-free case and is omitted from the figure. In this case, the crashed master may have already sent some commands or even the Commit Request to the switch.}
    \label{fig:rama-protocol-fault-case-2}
\end{figure}

When the master controller fails, the backup controllers will detect the failure (by timeout) and run a leader election algorithm to elect a new master for the switches.
Upon election, the new master must send a Role Request message to each switch, to register as the new master.
There are three main cases where the master controller can fail:

\begin{enumerate}
\item Before replicating the received event in the distributed log (Figure \ref{fig:rama-protocol-fault-case-1});
\item After replicating the event but before sending the Commit Request (Figure \ref{fig:rama-protocol-fault-case-2});
\item After sending the Commit Request message.
\end{enumerate}

In the first case, the master failed to replicate the received events to the shared log.
As slave controllers receive and buffer all events, no events are lost.
First, the new master must finish processing any events logged by the older master.
Note that events marked as processed have their resulting commands filtered.
This makes the new master reach the same internal state as the previous one before choosing the new order of events to append to the log (this is valid for all other fault cases).
The new elected master then appends the buffered events in order to the shared log and continues operation (feeding the new events to applications and sending commands to switches).

In the cases where the event was replicated in the log (cases 2 and 3), the master that crashed may or may not have issued the Commit Request message.
Therefore, the new master must carefully verify if the switch has processed everything it has received before re-sending the commands and the Commit Request message.
To guarantee ordering, OpenFlow provides a Barrier message, to which a switch can only reply after processing everything it has received before.
If a new master receives a Barrier Reply message without receiving a Commit Reply message (in form of \texttt{OFPT{\_}PACKET{\_}OUT}), it can safely assume that the switch did not receive nor execute a Commit Request for that event from the old master (case 2)\footnote{This relies on the FIFO properties of the controller-switch TCP connection.}.
Even if the old master sent all commands but did not send the Commit Request message, the bundle will never be committed and will eventually be discarded.
Therefore, the new master can safely resend the commands.
In case 3, since the old master sent the Commit Request before crashing, the new master will receive the confirmation that the switch processed the respective commands for that event and will not resend them (guaranteeing exactly-once semantics for commands).

%\vspace{.3cm}
\renewcommand{\arraystretch}{1.7}
\newcolumntype{A}{ >{\centering\arraybackslash} m{.28\linewidth} }
\newcolumntype{B}{ >{\centering\arraybackslash} m{.32\linewidth} }
\newcolumntype{C}{ >{\centering\arraybackslash} m{.32\linewidth} }
\begin{table*}[t!]
\begin{tabular}{ABC}
\hline
\vspace{-4px}\textbf{Property} & \vspace{-4px}\textbf{Ravana} & \vspace{-4px}\textbf{Rama}\\
\hline
\textit{At least once events} & Buffering and retransmission of switch events & Switches send events to every controller with role EQUAL\\
\hline
\textit{At most once events} & \multicolumn{2}{c}{Event IDs and filtering in the log}\\
\hline
\textit{Total event order} & \multicolumn{2}{c}{Master appends events to a shared log}\\
\hline
\textit{At least once commands} & RPC acknowledgments from switches & \multirow{2}{0.3\textwidth}[-0.1cm]{\centering{Bundle commit is known by every controller by piggybacking PacketOut in OpenFlow Bundle}}\\
\cline{1-2}
\textit{At most once commands} & Command IDs and filtering at switches & \\
\hline
\end{tabular}
%\caption{Consistency properties provided by Ravana and Rama}
\caption{How Rama and Ravana achieve the same consistency properties using different mechanisms}
\label{tab:rama-properties}
\end{table*}

%Note that to ensure that at least one controller will know that the switch completely executed the received commands it is important that the switch sends the Commit Reply message to all controllers (in our case, in the form of \texttt{OFPT{\_}PACKET{\_}OUT}).
%Imagine the other scenario where the switch only sends the Commit Reply message to the master controller.
%If the master controller fails after issuing the Commit Request to the switch but before receiving or replicating the reply to the other controllers, the new elected master would never know that the switch had committed the commands.
%As such, the inclusion of the \texttt{OFPT{\_}PACKET{\_}OUT} message to the bundles is the key to guarantee the level of consistency required.

\section{Correctness}

The Rama protocol we propose in this paper was designed to guarantee correctness of fault-tolerant SDN control.
We define correctness as in~\cite{katta2015}, where the authors introduce the concept of observational indistinguishability in the SDN context, defined as follows:

\emph{Observational indistinguishability:} If the trace of observations made by users in the fault-tolerant system is a possible trace in the fault-free system, then the fault-tolerant system is observationally indistinguishable from a fault-free system.

For observational observability, it is necessary to guarantee transactional semantics to the entire control loop, including (i) exactly-once event delivery, (ii) event ordering and processing, and (iii) exactly-once command execution.
In this section we summarize how the mechanisms employed by our protocol fulfil each of these necessary requirements.
For a brief comparison with Ravana, see Table \ref{tab:rama-properties}.

\textbf{Exactly once event processing:} events cannot be lost (processed \textit{at least once}) due to controller faults nor can they be processed repeatedly (they must be processed \textit{at most once}).
Contrary to Ravana, Rama does not need switches to buffer events neither that controllers acknowledge each received event to achieve \textit{at-least once event processing} semantics.
Instead, Rama relies on switches sending the generated events to \textit{all (f+1)} controllers (considering that the system tolerates up to \emph{f} crash faults) so that at least one will known about the event.
Upon receiving these events, the master replicates them in the shared log while the slaves add the events to a buffer.
As such, in case the master fails before replicating the events, the new elected master can append the buffered events to the log.
If the master fails after replicating the events, the slaves will filter the events in the buffer to avoid duplicate events in the log.
This ensures \textit{at-most once event processing} since the new master only processes each event in the log once.
Together, sending events to all controllers and filtering buffered events ensures \textit{exactly-once event processing}.

\textbf{Total event ordering:} to guarantee that all controller replicas reach the same internal state, they must process any sequence of events in the same order.
For this, both Rama and Ravana rely on a shared log across the controller replicas (implemented using the external coordination service) which allows the master to dictate the order of events to be followed by all replicas.
Even if the master fails, the new elected master always preserves the order of events in the log and can only append new events to it.

\textbf{Exactly once command execution:} for any given event received from one switch, the resulting series of commands sent by the controller are processed by the affected switches exactly \textit{once}.
Here, Ravana relies on switches acknowledging and buffering the commands received from controllers (to filter duplicates).
As this requires changes to the OpenFlow protocol and to switches, Rama relies on OpenFlow Bundles to guarantee transactional processing of commands.
Additionally, the Commit Reply message, which is triggered after the bundle finishes, is sent to \textit{all} controllers and thus acts as an acknowledgement that is independent of controller faults. If the master fails, the new master needs to know if it should resend the commands for the logged events or not.
A Packet Out message at the end of the bundle acts as a Commit Reply message to the slave controllers.
This way, upon becoming the new master, the controller replica has the required information to know if the switch processed the commands inside the bundle or not, without relying on the crashed master.
Furthermore, the new master sends a Barrier Request message to the switch.
Receiving the corresponding Barrier Reply message guarantees that neither the switch nor the link are slow (because a message was received and TCP maintains FIFO order) and thus there is no possibility of the Packet Out being delayed.
Therefore, the use of Bundles that include a Packet Out at the end, in addition to the Barrier message ensures that commands will be processed by the switches \textit{exactly-once}.

It is important to note that we also consider the case where switches fail.
However, this is not a special case of the protocol because it is already treated by the OpenFlow protocol under normal operation.
A switch failure will generate an event in the controller which will be delivered to applications, for them to act accordingly (e.g., re-route traffic around the failed switch).
A particularly relevant case is when a switch fails before sending the Commit Reply to the master and the slave controllers.
Importantly, this event does not result in transaction failure.
Since this is a normal event in SDN, the controller replicas simply mark pending events for the failed switch as processed and continue operation.

While we detail our reasoning as to why our protocol meets the correctness requirements of observational indistinguishability in SDN, modelling the Rama protocol and giving a formal proof is left as future work and out of the scope of this paper.

\section{Implementation}
\label{implementation}

\begin{figure}
  \centering
  \includegraphics[width=0.45\textwidth]{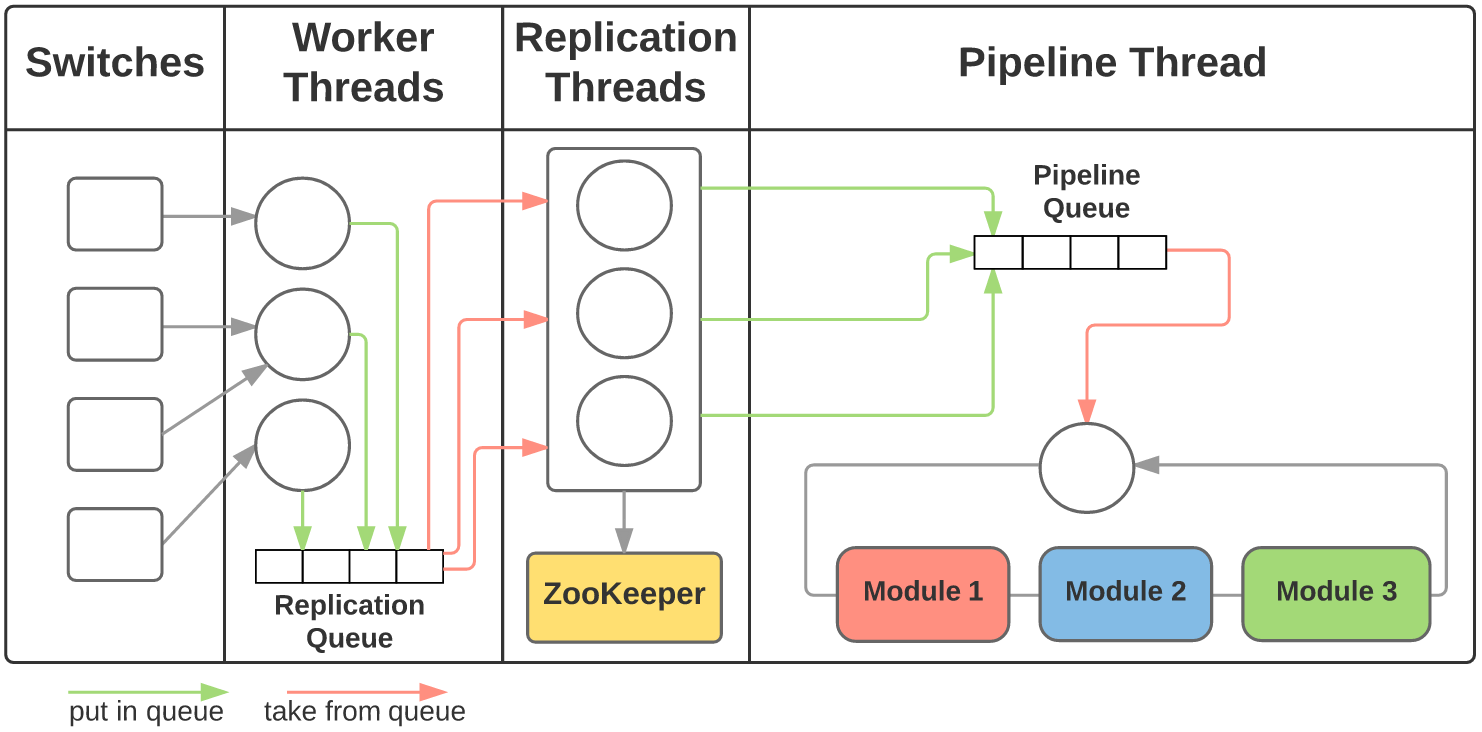}
  \caption{Rama thread architecture}
  \label{fig:rama-threads}
\end{figure}

We have built Rama on top of Floodlight~\cite{floodlight}.
For coordination, we opted for ZooKeeper~\cite{hunt2010}.
This service abstracts controllers from fault detection, leader election, and event transmission and storage (for controller recovery).
Rama introduces two main modules into Floodlight: the \textit{Event Replication} module (Section \ref{sec:event-replication}) and the \textit{Bundle Manager} module (Section \ref{sec:bundle-manager}).
Additionally, the Floodlight architecture was optimised for performance by introducing parallel network event collection and logging (Rama's multi-thread architecture is shown in Figure \ref{fig:rama-threads}) and by batching events (Section \ref{sec:event-batching}).
The multi-thread paralelism is introduced carefully, not to break TCP FIFO order of event processing, as will be explained next.

%\begin{figure}
%  \centering
%  \includegraphics[width=1\columnwidth]{figures/rama-architecture2.png}
%  \caption{Rama thread architecture with a single replication thread}
%  \label{fig:rama-threads}
%\end{figure}

In the original Floodlight, worker threads are used to collect network events and to process the modules pipeline (in Floodlight network applications are called ``modules'').
This design precludes event batching and other optimisations.
Ideally, we want to free the threads that collect network events as soon as possible so that they can keep collecting more events.
For this purpose, the worker threads' only job in Rama is to push events to the Replication Queue.
Events for a particular switch are collected always by the same thread (although each thread can be shared by several switches) and thus TCP FIFO order is guaranteed in the Replication Queue.
Next, the Rama runtime imposes a total order on the events by giving them a monotonically increasing ID.
As such, several Replication threads can then take events from this queue and execute the logic in the Event Replication module, which will send the events to ZooKeeper in batches, without breaking the required total order for correctness.
This technique is equivalent to Ravana's parallel event logging~\cite{katta2015}. 
When ZooKeeper replies to the request, the events are added to the Pipeline Queue to be processed by the Floodlight modules.
A single thread is used in this step, to guarantee the total order.
The slave replicas also follow the total order from the IDs assigned by the master.

One of our requirements was to make the control plane transparent for applications to execute unmodified.
The Event Replication module is transparent to other modules as it acts before the pipeline.
The modules will continue to receive events as usual in Floodlight and process them by changing their internal structures and sending commands to switches.
The process of sending messages inside OpenFlow Bundles as required by Rama is also made completely transparent to Floodlight modules, as will be explained in Section \ref{sec:bundle-manager}.

\subsection{Event Replication and ZK Manager}
\label{sec:event-replication}

The Event Replication module is the bridge between receiving events from the worker threads and pushing them into the pipeline queue to be processed by Floodlight modules.
Events are only added to the pipeline queue after being stored in ZooKeeper.
To separate tasks, Event Replication leverages on the ZK Manager, an auxiliary class that acts as ZooKeeper client (establishing connection, making requests and processing replies) and keeps state regarding the events (an event log and an event buffer in case of slaves) and switch leadership.
%Floodlight also has the notion of \textit{context} that is passed to the modules pipeline, but we can ignore it for now.
Event Replication and the ZK Manager work together to attain exactly-once event delivery and total order as follows.

When an event arrives at the Event Replication module, we check whether the controller is in master or slave mode.
In master mode the event is replicated in ZooKeeper and added to its in-memory log.
This log is a collection of \texttt{RamaEvent} objects which, apart from the switch information and message content, contains the unique event identifier explained before.
The events are replicated in ZooKeeper in batches (see Section \ref{sec:event-batching}), so each replication thread simply adds an event to the current batch and becomes free to process a new event.
Eventually the batch will be sent to ZooKeeper containing one or more events to be stored.
Upon receiving the reply, the events are pushed to the pipeline queue, ordered according to the identifier given by the master to guarantee total order.

In slave mode, the event is simply buffered in memory (to be used in the case where the master controller fails).
A special case is when the event received is the Packet Out that the master controller included in the bundle.
In this case, the slave marks that this switch already processed all commands for this event.
Slaves also keep an event log as the master, but only events that come from the master are added to it.
Events from the master arrive via \textit{watches} set in ZooKeeper nodes.
Slaves set watches and are notified when the master creates event nodes under that node. 
New events are added to the in memory log (so it is kept up-to-date with the log maintained by the master) and the events are added to the pipeline queue in the same way as in the master controller.
An important detail is that event identifiers are set by the master controller, and when slaves deserialize the data obtained from nodes stored in ZooKeeper they get the same \texttt{RamaEvent} objects created by the master.
Therefore, the events will be queued in the same order as they were in the master controller replica.

\subsection{Bundle Manager}
\label{sec:bundle-manager}
\begin{sloppypar}
The Bundle Manager module keeps state related to the open bundles for each switch (as result of an event) and is responsible for adding messages to the bundle, closing and committing it.
To guarantee transparency to applications, we modified the write method in \texttt{OFSwitch.java} (the class that is used by all modules to send commands to switches) to call the Bundle Manager.
This module will wrap the message sent by application modules in a \texttt{OFPT{\_}BUNDLE{\_}ADD\_MESSAGE} and send it to the switch.
This process is transparent because applications are unaware of the Bundle Manager module.
\end{sloppypar}
\begin{sloppypar}
In the end of the pipeline, the Bundle Manager module is thus called to prepare and commit the bundles containing the commands instructed by the modules as a response to this event.
Note that one event may cause modules to send commands to multiple switches, so in this step the Bundle Manager may send \texttt{OFPBCT\_COMMIT\_REQUEST} to one or more switches.
Before committing the bundle, the Bundle Manager also adds a \texttt{OFPT\_PACKET\_OUT} message to it, so that slave controllers will know if the commands for an event were committed or not in the switch (as explained in Section \ref{sec:ft_protocol}). 
This message will be received by the slave controllers as a \texttt{OFPT\_PACKET\_IN} message with the required information set by the master controller.
\end{sloppypar}

\subsection{Event batching}
\label{sec:event-batching}

Floodlight thread architecture was modified to allow event batching, for performance reasons.
Considering that ZooKeeper is running on a separated machine from the master controller replica, sending one event at a time to ZooKeeper would significantly degrade performance.
Therefore, the ZKManager groups events before sending them to ZooKeeper in batches.
Batches are sent to ZooKeeper using a special request called \texttt{multi}, which contains a list of operations to execute (e.g., create, delete, set data).
For event replication, the multi request will have a list with multiple create operations as parameter.
This request is sent after reaching the maximum configured amount of events (e.g., 1000) or some time after receiving the first event in the batch (e.g., 50ms).
This means that each event has a maximum delay bound (regarding event batching).
%Furthermore, to minimize the number of nodes created in ZooKeeper, each create operation will have data representing a list of events.
%This list size is bounded by the limitation imposed by the maximum allowable size of the data in each node, which is 1MB (1,048,576 bytes).
%This list size is bounded by the maximum allowable size of the data in each node, which is 1MB (1,048,576 bytes).

\section{Evaluation}
\label{evaluation}

In this section we evaluate Rama to understand its viability and the costs associated with the mechanisms used to achieve the desired consistency properties (without modifying the OpenFlow protocol or switches).

\subsection{Setup}

\begin{figure}[h]
  \centering
  \includegraphics[width=0.45\textwidth]{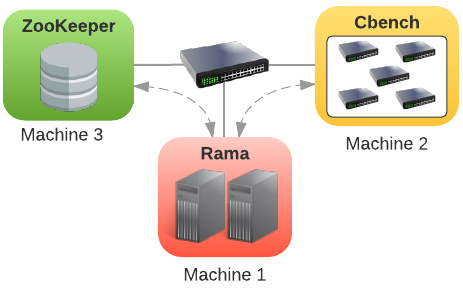}
  \caption{Experiment setup}
  \label{fig:setup}
\end{figure}

For the evaluation we used 3 machines connected to the same switch via 1Gbps links as shown in Figure \ref{fig:setup}.
Each machine has an Intel Xeon E5-2407 2.2GHz CPU and 32 GB (4x8GB) of memory.
Machine 1 runs one or more Rama instances, machine 2 runs ZooKeeper 3.4.8, and machine 3 runs Cbench to evaluate controller performance.
This setup tries to emulate a scenario similar to a real one with ZooKeeper on a different machine for fault-tolerance purposes, and Cbench on a different machine to include network latency. 

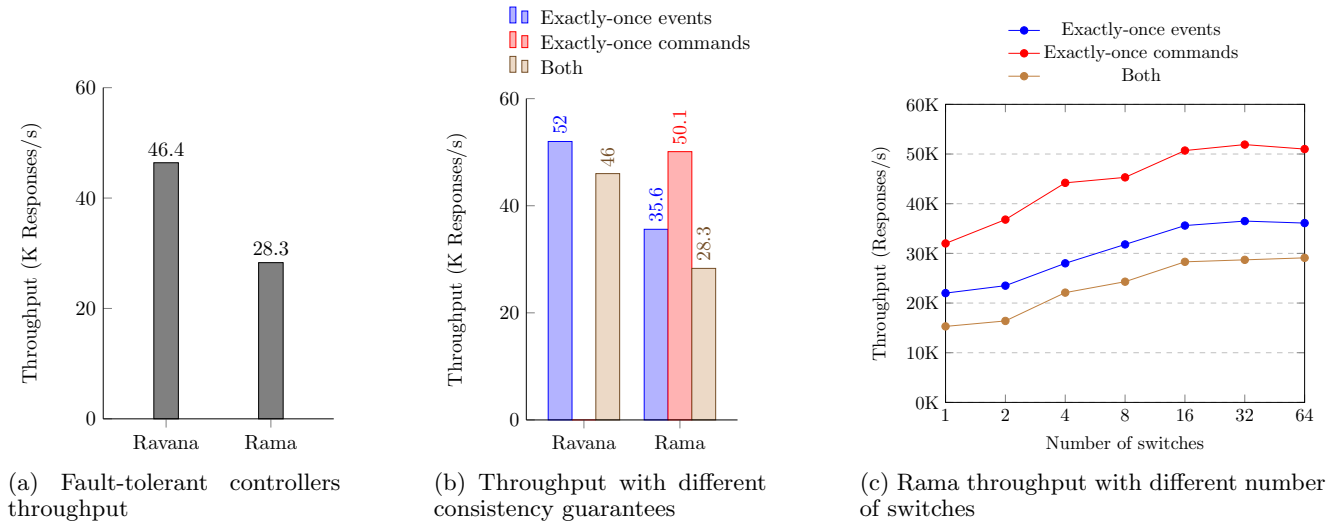
\begin{figure*}[h!]
	\centering
    \begin{subfigure}[t]{0.25\linewidth}
        \centering
        \resizebox{\columnwidth}{!}{
            \begin{tikzpicture}
				\begin{axis}[
						ybar=10pt,
						bar width=12pt,
						x=1.8cm, % distancia entre o centro das barras
						ymin=0,
						axis on top,
						ymax=60,
						ylabel=Throughput (K Responses/s),
						xtick=data,
						enlarge x limits=0.6, % distancia do 0 ate a primeira barra do x
                        symbolic x coords={Ravana,Rama},
						axis lines*=left,
						clip=false,
						transpose legend,
						legend style={draw=none,at={(0.5,1.3)},anchor=north},
                        nodes near coords,
                        cycle list name=black white,
                        every axis plot/.append style={fill=gray,no markers}
					]
					\addplot coordinates {(Ravana,46.4) (Rama,28.3)};
				\end{axis}
			\end{tikzpicture}
        }
        \caption{Fault-tolerant controllers throughput}
        \label{fig:controllers-throughput}
    \end{subfigure}
    \hspace{1cm}
    \begin{subfigure}[t]{0.25\linewidth}
        \centering
        \resizebox{\columnwidth}{!}{
            \begin{tikzpicture}
				\begin{axis}[
						ybar=0pt,
						bar width=12pt,
						x=1.7cm, % distancia entre o centro das barras
						ymin=0,
						axis on top,
						ymax=60,
						ylabel=Throughput (K Responses/s),
						xtick=data,
						enlarge x limits=0.6, % distancia do 0 ate a primeira barra do x
						symbolic x coords={Ravana,Rama},
						axis lines*=left,
						clip=false,
						transpose legend,
						legend style={draw=none,at={(0.5,1.3)},anchor=north, column sep=1.5mm},
                        legend cell align=left,
                        nodes near coords=\pgfmathfloatifflags{\pgfplotspointmeta}{0}{}{\pgfmathprintnumber{\pgfplotspointmeta}},
                        every node near coord/.append style={rotate=90, anchor=west},
					]
					\addplot coordinates {(Ravana,52) (Rama,35.6)};\label{legend-blue}
					\addplot coordinates {(Ravana,0) (Rama,50.1)};\label{legend-red}
					\addplot coordinates {(Ravana,46) (Rama,28.3)};\label{legend-brown}
					\legend{Exactly-once events, Exactly-once commands, Both}
				\end{axis}
			\end{tikzpicture}
        }
        \caption{Throughput with different consistency guarantees}
        \label{fig:ravana-vs-rama}
    \end{subfigure}
    \hspace{1cm}
    \begin{subfigure}[t]{0.35\linewidth}
        \centering
        \resizebox{\columnwidth}{!}{
            \begin{tikzpicture}
    \begin{axis}[
    xlabel={Number of switches},
    ylabel={Throughput (Responses/s)},
    xmin=1, xmax=7,
    ymin=0, ymax=60,
    xtick={1,2,3,4,5,6,7},
    xticklabels={1,2,4,8,16,32,64},
    ytick={0,10,20,30,40,50,60},
    yticklabels={0K,10K,20K,30K,40K,50K,60K},
    %legend pos=north west,
    legend style={draw=none,at={(0.5,1.3)},anchor=north},
    ymajorgrids=true,
    grid style=dashed,
    ]
    \addplot[
    color=blue,
    mark=*,
    ]
    coordinates {
    (1,22.0)(2,23.5)(3,28)(4,31.8)(5,35.6)(6,36.5)(7,36.1)
    };
    \addplot[
    color=red,
    mark=*,
    ]
    coordinates {
    (1,32.0)(2,36.8)(3,44.2)(4,45.3)(5,50.7)(6,51.9)(7,51.0)
    };
    \addplot[
    color=brown,
    mark=*,
    ]
    coordinates {
    (1,15.3)(2,16.4)(3,22.1)(4,24.3)(5,28.3)(6,28.7)(7,29.1)
    };
    \legend{Exactly-once events, Exactly-once commands, Both}
    \end{axis}
			\end{tikzpicture}
        }
        \caption{Rama throughput with different number of switches}
        \label{fig:rama-throughput-switches}
    \end{subfigure}
\caption{Throughput} 
\label{fig:rama-throughput}
\end{figure*}

\subsection{Rama performance}

We have compared the performance of Rama against Ravana~\cite{katta2015}.
Figure \ref{fig:controllers-throughput} shows the throughput for each controller (for Ravana we use the results reported in \cite{katta2015}, as its authors considered a similar setup).
For Rama measurements we run Cbench emulating 16 swit\-ches.

%Floodlight is optimized for performance achieving around 85K responses per second, with a configuration where %only core modules and a Hub application module are running.
%When we run Floodlight with a single worker thread so it comes down to Ryu's level (which has around 67K responses per second). 
Rama achieves a throughput close to 30K responses per second.
This figure is lower than Ravana's, as our solution incurs in higher costs compared to Ravana for the consistency guarantees provided.
The additional overhead is caused by two requirements of our protocol.
First, current switches' lack of mechanisms to allow temporary storage of OpenFlow events and commands require Rama to instruct switches to send all events to all replicas, increasing network overhead.
Second, the lack of acknowledgement messages in OpenFlow leads Rama to a more expensive solution -- bundles -- to achieve similar purposes.
The overhead introduced by these mechanisms is translated into reduced throughput when compared with Ravana.

In figure \ref{fig:ravana-vs-rama} we show, separately, throughput results considering the different levels of consistency provided by both Rama and Ravana.
The exactly-once events consistency level (\ref{legend-blue}) ensures that no events are lost and that controllers do not process repeated events.
Additionally, controllers must agree on a total order of events to be delivered to applications.
For the latter, both Rama and Ravana rely on ZooKeeper to build a shared log across controllers.
In our case, the master controller batches events in multiple requests to ZooKeeper, waits for replies, and orders the events before adding them to the Pipeline Queue.
Note that neither Rama nor Ravana wait for ZooKeeper to persistently store requests on disk (they both use ZooKeeper in-memory).
In our case, the multi-request is sent asynchronously (i.e., threads are freed to continue operation) and a callback function is registered. 
This function will be activated when ZooKeeper replies to our multi request and enqueues the logged events (in order) in the Pipeline Queue to be processed by the modules.
In Ravana the processing is equivalent.

The Exactly-once commands semantics (\ref{legend-red}) ensures that commands sent by controllers are not lost and that switches do not receive duplicate commands.
Ravana relies on switches to explicitly acknowledge each command and filter repeated ones.
For Rama, this includes maintaining state of all opened bundles for switches, and sending additional messages to the switches.
Instead of replying only with a Packet Out as in Floodlight, Rama must send messages to open the bundle, add the Packet Out to it, close the bundle and commit it.
To evaluate this case, we modified Cbench to make switches increase their counters only when they receive a Commit Request message from the controller.
This allows a fair evaluation of the performance of Rama in a real system -- indeed, in Rama a packet will only be forwarded after committing the bundle on the switch to guarantee consistent processing.

As show in Figure \ref{fig:ravana-vs-rama}, some guarantees are costlier to ensure than others\footnote{Note that we do not include the results from Exactly-once commands in Ravana as these are not available in~\cite{katta2015}. It is possible, however, to extrapolate that the results will be inline with the rest of the analysis.}.
For instance, the cost of providing Exactly-once events semantics is higher than Exactly-once commands semantics.
This result brings with it an important insight: the system bottleneck is the coordination service.
In other words, the additional mechanisms Rama uses to guarantee the desired consistency properties add overhead but, crucially, system performance is not limited by these mechanisms.

Figure \ref{fig:rama-throughput-switches} shows how maintaining multiple switch connections affects Rama throughput.
As switches send events at the highest possible rate, the throughput of the system saturates with around 16 switches.
Importantly, the throughput does not decrease with a higher number of switches.

%% NOVO
%\textbf{NOVO. }From the above comparison we can conclude that our initial prototype, although not quite good as Ravana, achieves an acceptable performance. The promising aspect of Rama is that the performance is more affected by the exactly-once events mechanisms than by the exactly-once commands mechanisms. Consequently, we are not limited by the overhead of having bundles in the protocol and have room to improve our performance (as can be seen by the throughput achieved by Ravana with exactly-once events mechanisms).

\subsection{Event batching}

Rama batches events to reduce the communication overhead of contacting ZooKeeper.
In practice, events are sent to ZooKeeper after reaching a configurable number of events in the batch (batching size) or after a configurable timeout (batching time).

\begin{figure}[t]
  \centering
  \resizebox{.75\linewidth}{!}{
  \begin{tikzpicture}
  \begin{axis}[
  xlabel={Batch size},
  ylabel={Throughput (Responses/s)},
  xmin=1, xmax=7,
  ymin=10, ymax=30,
  xtick={1,2,3,4,5,6,7},
  xticklabels={10,100,200,400,600,800,1000},
  ytick={15,20,25,30},
  yticklabels={15K,20K,25K,30K},
  ymajorgrids=true,
  grid style=dashed,
  ]
  \addplot[
  color=blue,
  mark=*,
  ]
  % 1   2   3   4   5   6   7 
  % 10,100,200,400,600,800,1000
  coordinates {
  (1,16.0)
  (2,16.9)
  (2.5,17.3)
  (3,17.9)
  (3.5,18.4)
  (4,19.1)
  (4.5,20.5)
  (5,22.6)
  (5.5,24)
  (6,24.9)
  (6.5,25.5)
  (7,28.3)
  };
  \end{axis}
  \end{tikzpicture}
  }
  \caption{Variation of Rama throughput with batch size}
  \label{fig:rama-batch-size}
  
\end{figure}
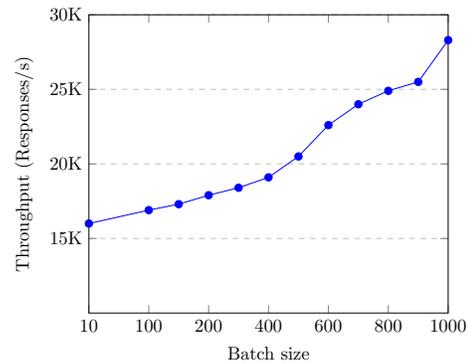

To evaluate batching we conducted a series of tests with different configurations to understand how the batching size and time affects Rama performance (Figure \ref{fig:rama-batch-size}).
Intuitively, a larger batching size will increase throughput, but as downside will also increase latency.
As batching size increases, throughput increases due to the reduction of RPC calls required to replicate events.

%\subsection{Latency}
%
%To measure Rama latency we run Cbench on the same machine as the master replica (to avoid network latency) on latency mode.
%Figure \ref{fig:rama-latency} shows that most events are processed in 9ms when using Rama with exactly-once events consistency and 12ms when running Rama with both consistency models.
%These results are inline with Ravana.

\subsection{Failover Time}

To measure the time for Rama to react to failures we use mininet~\cite{mininet}, OpenvSwitch~\cite{pfaff2015}, and iperf.
We setup a simple topology in Mininet with one switch and two hosts, one to act as iperf server and another as client.
We start the client and server in UDP mode, with the client generating 1 Mbit/sec for 10 seconds.
The switch connects to two Rama instances and sends all events to both controllers.
Each Rama instance is connected to the ZooKeeper server running on another machine (as before) with a negotiated session timeout of 500ms.
To make sure that no rules are installed on the switch -- so that events are sent to the controllers each time a packet arrives -- we run Rama with a module that only forwards packets (using Packet Out messages) without modifying the switch's tables.

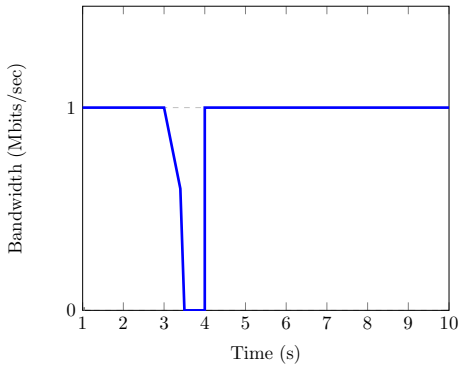
\begin{figure}[h!]
  \centering
  \resizebox{.75\linewidth}{!}{
    \begin{tikzpicture}
    \begin{axis}[
		xlabel={Time (s)},
		ylabel={Bandwidth (Mbits/sec)},
		xmin=1, xmax=10,
		ymin=0, ymax=1.5,
		xtick={1,2,3,4,5,6,7,8,9,10},
		%xticklabels from table={\datatable}{x},
		%xticklabels={1,2,3,4,5,6,7,8,9,10},
		ytick={0,1},
		%yticklabels={0K,10K,20K,30K,40K,50K},
		%yticklabels from table={\datatable}{y},
		%legend pos=north west,
		legend style={legend pos=south east},
		ymajorgrids=true,
		grid style=dashed,
		]
		\addplot[
		%smooth,
		color=blue,
		mark=none,
		line width=0.5mm
		]
		coordinates {
        (1,1)
        (3,1)
        (3.4,0.6)
        (3.5,0)
        (4,0)
        (4,1)
        (10,1)
		%(200,0)
		%(300,0)
		%(400,1/21)
		%(450,2/21)
		%(430,4/21)
		%(480,21/21)
		%(500,21/21)
		%(550,21/21)
		%(580,21/21)
		%(600,21/21)
		%(900,21/21)
		%(950,21/21)
		%(1000,21/21)
		};
        \draw[gray, dashed, thick] (3.4,0) -- (3.4,1.5);
        \draw[gray, dashed, thick] (4.0,0) -- (4.0,1.5);
		%\addplot[
		%smooth,
		%color=red,
		%mark=none,
		%line width=0.5mm
		%]
		%coordinates {
		%(40,0)
		%(45,0/21)
		%(50,0/21)
		%(55,0/21)
		%(60,0/21)
		%(65,1/21)
		%(70,6/21)
		%(75,8/21)
		%(80,13/21)
		%(85,21/21)
		%(90,21/21)
		%(95,21/21)
		%(100,21/21)
		%};
		\legend{}
    \end{axis}
  \end{tikzpicture}
  }
\caption{Rama failover time}
\label{fig:rama-failover}
\end{figure}

Figure \ref{fig:rama-failover} shows the reported bandwidth from the iperf server and indicates the time taken by Rama to react to failures. 
Namely, the slave replica takes around 550ms to react to faults.
This includes the time for: (a) ZooKeeper to detect the failure and notify the slave replicas (500ms); (b) electing a new leader for the swit\-ches; (c) the new leader to transition to master (finish processing logged events from the old master to reach the same internal state); (d) append buffered events to the log and start delivering unprocessed events in the log to applications so they start sending commands to the switches.
As is clear, the major factor affecting failover time is the time ZooKeeper needs to detect the failure of the master controller.

\subsection{Summary}

Rama comes close, but does not achieve the performance of Ravana.
This is due to the fact that our system incurs in higher costs. 
Rama requires more messages to be sent over the network and introduces new mechanisms, such as bundles, which increase the overhead of the solution in order to achieve the same properties as Ravana.
Despite the (relatively small) loss in performance, the value proposition of Rama of guaranteeing consistent command and event processing without requiring modifications to switches or to the OpenFlow protocol still makes it an effective enabler for immediate adoption of fault-tolerant SDN solutions.

%--------------------------------------------------------------------------------------------
\section{Related work}
\label{related}
%--------------------------------------------------------------------------------------------

\textbf{Consistent SDN.} Levin et al. \cite{levin2012} have explored the trade-offs of state distribution in a distributed control plane, motivating the importance of strong consistency in applications' performance.
On the one hand, view staleness affects the correct operation of applications, which may lead to poor network performance.
On the other, applications need to be more complex in order to be aware of possible network inconsistencies.

Having a strongly consistent network view across the controllers may be critical to the operation of some applications (e.g., load balancing) in terms of correctness and performance \cite{levin2012}.
However, as noted in the CAP theorem, a system can not provide availability while also achieving strong consistency in the presence of network partitions.
Because of this, fault-tolerant and distributed SDN architectures must use techniques to explicitly handle partitions in order to optimize consistency and availability (and thus achieving a tradeoff between them)~\cite{brewer2012}.

Part of the strong consistency in the controllers comes from a consistent packet processing (i.e., packets received from switches). OF.CPP \cite{perevsini2013} explores the consistency and performance problems associated with packet processing at the controller and proposes the use of transactional semantics to solve them.
These semantics are achieved by using multi-commit transactions, where each event is a sub transaction, which can commit or abort, of the related packet (the main transaction).
However, this transactional semantics in packet processing is not enough: controllers should also coordinate to guarantee the same semantics in the switches' state. 
Specifically, the commands sent by the controllers should be processed exactly once by the corresponding switches -- a problem our work addresses.

\textbf{Consistent network updates.} The concepts of per-packet and per-flow consistency in SDN were introduced in \cite{reitblatt2011} to provide a useful abstraction for applications: consistent network updates.
With consistent updates, packets or flows in flight are processed exclusively by the old or by the new network policy (never a mix of both).
For example, with per-packet consistency,  every packet traversing the network is processed by exactly one consistent global network configuration.
The authors extend this work in \cite{reitblatt2012} and implement Kinetic, which runs on top of NOX \cite{gude2008} to offer these abstractions in a control plane to be used by applications.
The main mechanism used to guarantee consistent network updates is the use of a two-phase protocol to update the rules on the switches.
First, the new configuration is installed in an unobservable way (no packets go through these rules yet).
Afterwards, the switch's ingress ports are updated one-by-one to stamp packets with a new version number (using VLAN tags).
Only packets with the new version number are processed by the new rules.

In \cite{canini2013}, Canini et al. extend Kinetic to a distributed control plane and formalize the notion of fault-tolerant policy composition.
Their algorithm also requires a bounded number of tags, regardless of the number of installed updates, as opposed to the unbounded number of tags in \cite{reitblatt2012}.

This class of proposals addresses consistent network updates, which is an orthogonal problem to the one addressed here.

\textbf{Fault-tolerance in SDN.} Botelho et al.~\cite{botelho2014} and Katta et al.~\cite{katta2015} both address fault tolerance in the control plane while achieving strong consistency.
In~\cite{botelho2014} the authors proposes SMaRtLight, a fault-tolerant controller architecture for SDN. 
Their architecture uses a hybrid replication approach: passive replication in the controllers (one primary and multiple backups) and active replication in an external distributed data store, to achieve durability and strong consistency.
The controllers are coordinated through the data store and ca\-ching mechanisms are employed to achieve acceptable performance.
In~\cite{botelho2016} the authors extend their solution to a distributed deployment.
In contrast to our solution, SMaRtLight requires applications to be modified to use the data store directly.
More importantly, the solution does not consider the consistency if switch state in the system model.
Ravana~\cite{katta2015} was the first fault-tolerant controller that integrates switches into the problem.
The techniques proposed by its authors guarantee correctness of event processing and command execution in SDN.
The main differentiating factor of our work to Ravana is that our solution does not require changes to the OpenFlow protocol nor to switches.

\textbf{Distributed SDN controllers}. The need for scalability and dependability has been a motivating factor for distribution and fault-tolerance in SDN control planes.
Onix~\cite{koponen2010}, the first distributed, dependable, production-level solution considered these problems from the outset.
As the choice of the ``right'' consistency model was perceived as fundamental by its authors, Onix offered two data stores to maintain the network state: an eventually consistent and a strong consistent option.
ONOS~\cite{berde2014} is an open-source solution that shares with Onix the fact that controller state is stored in an external consistent data-store.
Both approaches ensure a consistent ordering of events between controller replicas, but they do not include switch state and hence can lead to the network anomalies of traditional replication solutions. 

\textbf{Traditional fault-tolerance techniques.} Viewstamped Replication~\cite{oki1988}, Paxos~\cite{lamport1998}, and Raft~\cite{ongaro2014} are well-known distributed consensus protocols used for replication of state machines in client-server models.
None of these widely-used protocols is directly applicable in the context of SDN, where to guarantee correctness it is necessary not only to have consistent controller state, but also switch state.

\section{Conclusions}

In a fault-tolerant SDN, maintaining consistent controller state is not enough to achieve correctness.
Unlike traditional distributed systems, in SDN it is necessary to consistently handle switch state to avoid loss or repetition of commands and events under controller failures.
To address these challenges we propose Rama, a consistent and fault-tolerant SDN controller that handles the entire event processing cycle transactionally.

Rama differs from the existing alternative, Ravana ~\cite{katta2015}, by not requiring modifications to the OpenFlow protocol nor to switches.
This comes at a cost, as the techniques introduced in Rama incur in a higher overhead when compared to Ravana.
As the overhead leads to a relatively modest decrease in performance, we expect, in practice, this to be compensated by the fact that our solution is immediately deployable.
We make our software available open source\footnote{https://github.com/fvramos/rama} to further foster adoption of fault-tolerant SDN.

As for future work, besides devising a formal proof on the consistency guarantees Rama provides, we plan to address correctness in distributed SDN deployments and to consider richer fault models. 

%ACKNOWLEDGMENTS are optional
%\section{Acknowledgments}
%This section is optional; it is a location for you
%to acknowledge grants, funding, editing assistance and
%what have you.  In the present case, for example, the
%authors would like to thank Gerald Murray of ACM for
%his help in codifying this \textit{Author's Guide}
%and the \textbf{.cls} and \textbf{.tex} files that it describes.

\bibliographystyle{abbrv}
\bibliography{refs}  % sigproc.bib is the name of the Bibliography in this case

\end{document}